Title: Two-electron F' centers in alkali halides: A negative-U analysis

Authors: Mladen Georgiev (Institute of Solid State Physics, Bulgarian Academy of Sciences, Sofia, Bulgaria)

The existence of bound excited states of two-electron centers loosely trapped in an anion vacancy, the F' centers, is a long-standing problem of color center physics. Optical absorption bands attributed to F' centers in NaI and NaBr as observed by *Baldacchini et al*. seem to confirm the existence of F' bound singlet states. Nevertheless, a paper by *Zhang et al.* published not long ago reports extended ion calculations predicting bound excited states for F' centers in NaI alone while the evidence for NaBr is only marginal. For the first time we apply the negative-U mechanism in order to see whether bound excited F' states do not arise by virtue of the lattice polarization produced by the two trapped charge carriers. The result is positive for NaI, and less so for NaBr, in agreement with *Zhang et al.* Other hosts, such as CsI and RbI also show a tendency towards sustaining F' bound excited states. However, the lack of entry data for LiI makes it presently too hard to obtain a complete negative-U picture for the iodides.

## 1. Introduction

### 1.1. Rationale

The F' center is undoubtedly one of the most appealing species among the variety of color centers in the alkali halides.[1] Certainly, the scientific interest in these “simple” species has not ceased lately.[2] This is a two-electron center trapped at an anion vacancy, a cavity-localized electronic dimer. In most crystalline hosts, such as KCl, NaCl, and KBr, the role of the anion vacancy in forming the bond that holds the two electrons together against the Coulomb repulsion seems to be decisive: It is the depth of the positive potential well that holds two electrons at a time. However, in other less numerous cases, such as NaBr, NaI, and possibly LiCl, the lattice polarization appears to be the major factor that helps confine the electrons inside the same well. F' centers in salts of the latter group have shown certain peculiar behavior, such as the occurrence of a strong bell-shaped F' absorption band on the short wavelength side of the F band.[3] For comparison, the F’ band in saltts of the former group is structureless and extends to the red of the F band. For the latter group a conceivable mechanism to secure an effective F' binding is Anderson’s negative-U.[4] If that is the case, the F' center of the latter category will come as the primary species in the sequence of electron dimers from amorphous semiconductors to high-$T_c$ superconductors. Then, studying F' centers in alkali halides may give the key to understanding electronic molecules wherever they appear.

It is the purpose of a current investigation to see whether the peculiar properties of F' centers in sodium salts can be explained by the negative-U mechanism. In what follows we first outline the basic premises of that mechanism, then check it against what is presently known of the less common F' centers.

## 1.2. Brief survey of F' work after 1988

A detailed account of the work on F' centers in alkali halides prior to 1988 can be found elsewhere.[1] Lately, *Zhang et al.* have presented the results of their calculations of the F' thermal and optical ionization energies, as well as of a search for excited F' singlet and triplet states in several host halides..[2] In effect the authors have made use of the extended-ion method appropriate for studying excited-states in ionic crystals which has been applied earlier to the F' center problem by *Strozier and Dick*.[5]

In view of the importance of *Zhang et al.*'s analysis and its close relationship to the present work, it seems appropriate to devote some more space to recapitulating though briefly the essentials of their method. The purpose is to make transparent similarities and differences. The authors have used floating 1s Gaussians $G(\alpha_{k,i},\xi_{k,i};\mathbf{r}_i)$ orthogonalized to the occupied core states $\chi_{\gamma,\lambda}(\mathbf{r}_I)$ where $\xi_{k,i}$ are the lattice configurations corresponding to the different electronic states. The Hamiltonian is $H = H_1 + H_2 + H_{12}$ where

$$H_i = -(\hbar^2/2m)\Delta_i + V_{PI}(r_i) + [V(r_i) - V_{PI}(r_i)] \text{ for } i = 1,2 \qquad (1)$$

is the one-electron Hamiltonian in which $V_{PI}(\mathbf{r}_i)$ is the long-range point-ion potential, while $V(\mathbf{r}_i)$ contains short range Coulomb and exchange terms. The *e-e* interaction term is

$$H_{12} = e^2 / r_{12} \qquad (2)$$

where $r_{12} = |r_1 - r_2|$ reflects the requirement that the trial wave function take account of the *e-e* correlations which is partially met by a sum of product functions by Gaussians. Accordingly, the two-electron F' wavefunction is taken as a LC of basis set product functions:

$$\Phi(r_1,r_2) = \sum_k C_k\, \phi_{k,1}(\mathbf{r}_1)\, \phi_{k,2}(\mathbf{r}_2)$$

$$\phi_{k,i}(r_i) = G_{k,i}(r_i) - \sum_{\gamma\lambda_} < \chi_{\gamma,\lambda} \mid G_{k,i} > \chi_{\gamma,\lambda}(r_I) \quad (i = 1,2)$$

$$G_{k,i}(\mathbf{r}_i) = N_{k,i} \exp( -\alpha_{k,I} |\mathbf{r}_I - \xi_{k,i}|^2 ) \qquad (3)$$

The lattice energy is considered consisting of two parts, Coulombic and repulsive. The former is the electrostatic energy of an infinite lattice of point ions, the latter is taken in the Born-Mayer form. For describing the polarization energy, the lattice is divided into two subregions: I (short ranged in the immediate neighborhood of the defect) and II (long ranged far from the defect). The short range polarization is accounted for using Mott-Littleton's discrete method, the long range polarization is dealt with by means of a continuum method due to Hardy and Lidiard. The resulting energy functional $E = E_{el} + E_{latt} + E_{polI} + E_{polII}$ is minimized subsequently in $\alpha$ and in $\xi$.

The optical energy is defined as

$$E_{op}(F') = E_g(F^*) - E_g(F') - \chi \qquad (4)$$

where both the F and F' ground state energy components are taken at the lattice configuration appropriate for the F' ground state, $\chi$ is the electron affinity of the crystal. The thermal ionization energy is:

$$E_{th}(F') = E_g(F) - E_g(F') - \chi \qquad (5)$$

now each of the energy components being taken at the equilibrium configuration of its respective electronic state.

The calculated optical and thermal energies of F' centers in NaCl, KCl, NaBr and NaI compare favorably with the experimental data. Most remarkably, they show that the F' potential deepens as one goes from NaCl to NaI and predict the existence of F' singlet and triplet excited states for NaI, while for NaBr the suggestion is for a marginal existence. The calculated optical transition energy of the F' ground-state singlet to the F' excited-state singlet (1.9 eV) and the ionization energy of the F' excited-state triplet (1.3 eV) in NaI appear to be similar to the observed energies from the respective optical absorption bands, 2.0 eV and 1.45 eV, respectively. See reference [1] for an outlook of the preceding experimental work.

I remember having met both *Song* and *Leung* at a conference. *Leung* asked me which F' potential was best. “The negative-U”, I replied without any hessitation. *Leung* exclaimed in surprise, but I now think he did so more because of my frankness rather than because of the essence of my statement. I believe they themselves have not been far from the idea at that time (1989 to my best memory).

## 1. Negative-U mechanism

### 2.1. Theoretical background

Before all, we remind of the unusual optical properties of F and F' centers in crystals with a small cation-to-anion radius ratio. They seem to have a stable excited state as optical transitions from the ground state lead to a bell-shaped band shifted on the violet side of the F band. We are thereby stimulated to describe the singlet F' center in terms of a second-quantization atomic-orbital model in site-representation based on an extended Hubbard Hamiltonian:

$$H = \sum_{(ij)} t_{ij}\, a_{i\sigma}^{\dagger} a_{j\sigma} + \sum_n E_n a_{n\sigma}^{\dagger} a_{n\sigma} + \sum_n U_n\, a_{n+\sigma}^{\dagger} a_{n+\sigma} a_{n-\sigma}^{\dagger} a_{n-\sigma} +$$

$$\tfrac{1}{2}\sum_n K_n\, q_n^2 - 2\sum_n G_n\, q_n a_{n\sigma}^{\dagger} a_{n\sigma}$$

$$= t_{F'} \sum_{(ij)} a_{i\sigma}^{\dagger} a_{j\sigma} + E_{F'} \sum_n a_{n\sigma}^{\dagger} a_{n\sigma} + U \sum_n a_{n+\sigma}^{\dagger} a_{n+\sigma} a_{n-\sigma}^{\dagger} a_{n-\sigma} +$$

$$\tfrac{1}{2}Kq^2 - 2G\, q \sum_n a_{n\sigma}^{\dagger} a_{n\sigma} \qquad (6)$$

where $t_{F'}$ is the electron hopping energy between nearest-neighbor sites (ij) (accounting for the possibility that the F' electron hops to a neighboring F center), $E_{F'}$ is the binding energy of an electron in the field of the anion vacancy, U is the electron-electron (*e-e*) correlation energy, G is the spin-independent electron-phonon coupling constant. $a_{n\sigma}^{\dagger}$ ($a_{n\sigma}$) are electron creation (annihilation) operators, *q* is the phonon coordinate. The factor 2 to G arises, since there are two F' electrons. We consider a system of F' centers and F centers at sites n,m (in a simplified approach we neglect the empty anion vacancies, the $\alpha$-centers).

Assuming a classical lattice, minimizing H in *q* we get two extrema: one at $q_0 = 0$ and another one at $q_l = 2(G/K) \sum_{n\sigma} a_{n\sigma}^{\dagger} a_{n\sigma}$. Inserting $q_l$ back and using $q_l^2 = 4(G/K)^2 \times \sum_{n\sigma_} a_{n+\sigma}^{\dagger} a_{n+\sigma} a_{n-\sigma}^{\dagger} a_{n-\sigma}$ because of Pauli's exclusion principle, we eliminate the phonon coordinate to get

$$H_{F'} = t'_{F'} \sum_{(ij)} a_{i\sigma}^{\dagger} a_{j\sigma} + E'_{F'} \sum_{n\sigma} a_{n\sigma}^{\dagger} a_{n\sigma} + (U - 4E_{LR}) \sum_{n\sigma} a_{n+\sigma}^{\dagger} a_{n+\sigma} a_{n-\sigma}^{\dagger} a_{n-\sigma} \quad (7)$$

Similar arguments lead to the following Hamiltonian of an F center:

$$H_F = t'_F \sum_{(ij)} a_i^{\dagger} a_j + E'_F \sum_n a_n^{\dagger} a_n - E_{LR} \sum_n a_n^{\dagger} a_n . \quad (8)$$

The spin variable is now irrelevant. The primed quantities in eqs. (7) and (8) are those renormalized by the electron-phonon coupling. Note that in the latter F case the lattice relaxation energy $E_{LR} = G^2/2K$ enters one-fold. In the former F' case, the four-fold lattice-relaxation energy $4E_{LR} = 2G^2/2K$, lowers the Coulomb repulsion energy and can even overcompensate it to make $U_{eff} = 4E_{LR} - U > 0$. Excluding the phonon coordinate q is thus seen to result in the appearance of a negative-U center (if $U_{eff} > 0$) identifying the F' center as an Anderson bipolaron bound to the field of an anion vacancy.

In order to calculate $E'_{F'}$, the quantum mechanical electron binding energy in the field of an anion vacancy in the presence of both *e-e* correlations and lattice relaxation effects, we solve for the respective Schrödinger equation with a square-box potential where most of the Franck-Condon states in absorption are situated:

$$[\sum_{i=1,2}(\hbar^2/2m_i)\Delta_i + V(\mathbf{r})]\psi(\mathbf{r}) = E_{F'}\psi(\mathbf{r}),\ V(\mathbf{r}) = \begin{cases} V_0, & r \leq a \\ 0, & r > a \end{cases} \quad (9)$$

where $V_0 = V_M - \chi + 4E_{LR} - U$. Solving for s-states, we get [6]

$$E_{F'} = -(\hbar y/a)^2/2m \quad (10)$$

where y is the solution of the transcendent equations:

$$y = -x\cot\tan x \quad (11)$$

$$x^2 + y^2 = 2mV_0(a/\hbar)^2 \equiv R^2 \quad (12)$$

We set $m_i = m$ for the electron mass and $R = (a/\hbar)\sqrt{(2mV_0)}$. At $\sqrt{(2mV_0)}\,[2a/\pi\hbar] < 1$, i.e. $(2/\pi)R < 1$ (range I) there are no bound states. At $1 \leq \sqrt{(2mV_0)}\,[2a/\pi\hbar] < 3$, i.e. $1 \leq (2/\pi)R < 3$ (range II) there is a single bound state and there are further bound states at $3 \leq \sqrt{(2mV_0)}\,[2a/\pi\hbar]$, i.e. $3 \leq (2/\pi)R$ (range III).

Solving for x and rationalizing, one finds

$$x/\sin x = \pm R, \text{ or } (2/\pi)R = \pm(2/\pi)(x/\sin x). \qquad (13)$$

Range borders $(2/\pi)R$ for various alkali halides are delineated in the underlying Tables. Most halides are seen falling in the single-bound-state range II, though the iodides and bromides fall in close to the borderline with range III. Three compounds of the latter categories: RbI, NaI, and possibly CsI eventually cross over the range III borderline and are thus predicted to having more than one bound state.

Vanishing $x = 0$ which require $R = 1$ to solve the eigenvalue equations fall within range I ($(2/\pi)R < 1$) of no bound states. Other solutions are calculated numerically for alkali halides which make it possible to identify energy levels associated with F' bound states. The F' bound state energies are calculated from the eigenvalue formula:

$$E_{F'n} = -(\hbar^2/2ma^2)y_n^2 \qquad (14)$$

where $y_n$ is the corresponding root of the eigenvalue equation. These energy levels lie below the bottom of the conduction band at $E = 0$. For calculating the electronic energies, a tentative value of $m = 0.5\, m_e$ ($m_e$ is the free electron mass) is used for the electron effective mass. The corresponding eigenstates are interpreted as F' bound states, ground and excited states active in absorption, that is, Frank-Condon states. The optical absorption energies are obtained as energy differences between the bound-state energies. This holds good if the Coulomb repulsion between the F' electrons is of a comparable magnitude in ground state and in the excited states, as well as the lattice relaxation energy.

Deriving the *e-e* correlation energy is somewhat easier. We see the correlation in that the two s-electrons move along a common circular orbital at a constant angle $(r_1,r_2) = \pi$. Therefore:

$$U = {}_0\int^{\infty} A_1\exp(-\mathbf{k}\mathbf{r}_1)[e^2/\varepsilon|\mathbf{r}_1-\mathbf{r}_2|]A_2\exp(-\mathbf{k}\mathbf{r}_2)d\mathbf{r}_1d\mathbf{r}_2 \,/\, {}_0\int^{\infty} A_1\exp(-\mathbf{k}\mathbf{r}_1)A_2\exp(-\mathbf{k}\mathbf{r}_2)d\mathbf{r}_1d\mathbf{r}_2 \qquad (15)$$

In view of the assumed correlation, $|r_1-r_2| = \sqrt{(r_1^2 + r_2^2 + 2r_1r_2)} = (r_1 + r_2) \sim 2a$, where $a$ is the effective s-electron orbital radius. As a result we get:

$$U \approx e^2/2\varepsilon a \qquad (16)$$

which, even though somewhat simplified, is expected to give the correct order of magnitude. Finally the electron-phonon linear-coupling coefficient is G = $V_M/a$ wherefrom

$$E_{LR} = G^2/2K = (V_M/a)^2/2K \qquad (17)$$

where $V_M$ is Madelung's potential, $K = M\omega^2$ is the spring constant of the breathing mode oscillator of mass $M = 6M_{metal}$.

## 2.2. Numerical calculations

The orbital radius *a* should be close to the effective vacancy radius $r_0$, often referred to above. Paradoxically, one expects smaller values of *a* to bring about higher binding energies and vice versa. Smaller values of *a* and respective higher binding energies should be associated with F' centers, due to the extra negative charge which pulls the neighboring cations closer to the vacancy center and reduce its radius. An inward relaxation of the nearest-neighbor cation 'sphere' towards the anion vacancy center is also the fingerprint of the attractive interaction associated with the electron-lattice coupling which overweighs the Coulomb repulsion and leads to $U_{eff} > 0$. Nevertheless, F' binding energies in most alkali halides are inferior to F center binding energies: As a rule F bands are known lying on the violet side of F' band plateaus. Exceptions are provided by crystals with small cation-to-anion radius ratios, $r_{cat}/r_{an} < 0.5$, for which the above geometric considerations may apply. In our opinion, however, the reasons for the controversy may rather be in the electric parameters, such as Madelung's potential, dielectric constant, etc.

Semicontinuum parameters for F centers entering in the respective formulas are listed in Table 1. The electron affinity is taken to be $\chi = 0.5$ and the electron effective mass is taken to be $m = 0.5\ m_e$ for all the crystalline hosts. Both are compromise choices based on data from Fowler's book. We do not know the exact nature of the vibrational mode coupled to the F' center. For this reason we tabulate experimental vibrational frequency data for two modes, the local $A_{1g}$ breathing mode (BMO) from Raman data coupled to the F center in ground state (frequency $\omega_{Fg}$) and the longitudinal-optic (LO) mode of the crystalline host (frequency $\omega_{LO}$).[7-10] As a rule, estimates are carried out at either frequency.

It should be mentioned that whlle the BMO concept has gained legitimacy over the years as regards the one-electron single-vacancy F color center,[7] further motivation will be needed to justify applying similar arguments to the two-electron single-vacancy F' center. Nevertheless, in view of retaining the configurational structure the F' coupled mode may be akin by symmetry to the F center breathing mode though vibrating at a somewhat higher frequency. (A frequency hardening may result from the stiffening of the F' environment relative to the F center because of the extra electron.) In any event, we introduce an alternative higher frequency breathing mode to couple to the F' center taking $\omega_{LO}$ as an attempt frequency alternative to $\omega_{Fg}$.

Negative-U parameters for F' centers are listed in Table 2. There are two sets of data calculated at $\omega_{Fg} \equiv \omega_{BMO}$ and $\omega_{LO}$, respectively. For obtaining the stiffness $K = M\omega^2$ which enters into the Jahn-Teller energy $E_{JT}$ in either case the oscillator mass is estimated as $M = M_{BMO} = 6M_{metal}$.[1] Table 2 also presents range borders $(2/\pi)R$ and experimental F' optical energies. The latter are F' band threshold energies ($E_{F'edge}$) from the near-infrared edges of the F' bands and F' peak energies ($E_{F'peak}$) from the observed positions of the F' band peaks (usually broad), as listed separately.

Roots $x_n$ and $y_n$, and F' bound-state energies $E_{F'n}$ calculated for five alkali halides are shown in Table 3. The first and third columns therein present $x_n$ and $y_n$ derived using $\omega_{BMO}$, while the second and fourth columns show results obtained using $\omega_{LO}$. For some crystalline hosts we present more than one set of $\omega_{BMO}$ based calculations reflecting different literature data for the frequency coupled to the F ground state, as explained in the footnote to Table 1. If F' centers in a crystalline host do have excited bound states, then the corresponding peak energies should be interpreted as separations between their respective bound states. Optical transitions between S states being forbidden by parity, the calculated optical energies should lie in-between the experimental $E_{F'edge}$ and $E_{F'peak}$ which is roughly observed. It is implied that once an S bound excited state is identified, a lower-lying P state accompanies it close in energy, as it does for F centers.

Table 3 calculations have been made for F' systems whose energies lie either above the bound-excited-state borderline or are less than 2.5 percent below it to allow for possible computational underestimates. The tabulated estimates of the F' bound excited state energies are sensitive to the choice of a coupled vibrational frequency. Most of the alkali halide hosts sustain a single F' bound S state (range II in Table 2). The exceptions are mentioned below: The NaI data (range III) reveal a clear trend towards the formation of an F' excited bound state making the NaI F' center perhaps the strongest candidate thus far for a negative-U identity. We also see CsI to form a bound excited state at the LO frequency with an optical excitation energy between the experimental Table 2 values. Another clear host that may sustain F' bound excited states is RbI. Eventually so may KBr, and possibly RbBr, both being situated very closely below the two-state borderline. However, the F' center in NaBr has clearly no bound excited S states, though it may have lower-symmetry bound excited P states not covered by the present analysis.

## 3. Discussion

The experimental criterion $r_+/r_- < 0.5$ has been introduced empirically to distinguish between “unconventional” and “normal” F' centers. It signifies that a cation radius smaller than half the anion radius should be essential for determining the optical properties of F' centers. Perhaps it would be hard to see just how meeting this condition would lead to deepening the electronic potential at the anion vacancy. One way may be considering the local lattice relaxation energy and indeed that energy is given by $E_{LR} = G^2/2K$ where G is the appropriate coupling constant. Smaller cation radii would imply a softer anion lattice which would elevate the coupling at the anion vacancy. If so, the ionic criterion would imply an useful interrelation between ionic radii and electron-phonon coupling strength.

From Table 1 we see that while NaI complies fully with the ionic radii criterion, so do most of the Li salts. At the same time, NaBr's cation is 2.5 percent oversized, NaCl's is 5.8 percent oversized, and LiF's is only 0.7 percent oversized. In view of the lack of any definite data on unconventional F' centers in the oversized salts let alone on the Li salts, we conclude that the ionic criterion does not seem to suffice.

The semicontinuum negative-U model employed presently predicts a single bound state for most of the two-electron centers in alkali halides which seems to give credit to the theory. Some of the salts fall not too far on the left side of the $(2/\pi)R = 3$ borderline, which separates F' centers with a single bound (ground) state from centers with at least one bound excited state. For NaI the prediction is for a bound excited state ($(2/\pi)R>3$). Another host which may have a bound excited state is RbI but this has not been confirmed experimentally. For NaBr the conclusion is less than affirmative. There are a few other salts too close to the borderline at $(2/\pi)R = 3$. Unfortunately, the raw data on the Li salts are only scarce which does not allow for the drawing of any safe conclusions.

It is also noteworthy that the computed two sets of differences in Table 2 of the F' bound state energies for NaI are 2.10 eV and 1.61 eV, respectively. These compare favorably with the experimental F' peak at 2.4 eV and the F' edge at 1.9 eV. Clearly, this once again gives credit to the negative-U assignment to the F' center in NaI. However, the result by *Zhang et al.*[2] of a deepening F' potential in the sequence NaCl→NaBr→NaI is not confirmed by the present negative-U analysis (cf. Table 2), the reason being that the negative-U effect depends on the product $V_0 \times a^2$ (rather than on $V_0$ alone) where the interionic separation $a$ drops steadily along that sequence.

The effect of the vacancy field on the F' bound state energies has yet to be established. Of the phenomenological potentials often employed in color center theory for quick albeit meaningful estimates of bound state energies and thereby of optical spectra, the semi-continuum models lie perhaps closest to the present negative-U potentials except for the lattice relaxation energies not accounted for semi-continually because of the traditional assumption that the F' bands are insensitive to the coupled phonon frequencies and thereby to the temperature.[1] Clearly the temperature independency has proved questionable in the light of observed F and F' bands in sodium salts with small cation-to-anion radii ratios. Apart from the relaxation effects, negative-U and in-box semi-continuum energy spectra are akin to each other and are therefore subject to superposition and interference.

Both semi-continuum and negative-U potentials comprise considerable spherical-box parts, so the vacancy-field problem reduces to the relative contribution of the Coulomb tail to the former part. This tail is the less significant the wider and deeper the box. Thus we are again led to conclude that the ionic radii do play a role. Of the alkali halide hosts listed in Table 1 the iodides provide wider boxes. However, excited bound states are more likely to occur in spherical box potentials deeper in magnitude, such as provided by the fluorides.

Another problem worth attention is the electron hopping $t'_{F'}$ between neighboring sites at higher color center densities. Due to electron-phonon coupling, the electron operators renormalize by the amount of the vibrational overlap at the F' center sites. For all diagonal matrix elements this gives 1, while for the off-diagonal elements the result is the overlap of the displaced oscillator wave functions at neighboring sites. In effect, the electron hopping energy renormalizes to a magnitude exponentially reduced relative to its bare value:

$$t'_{F'} = t_{F'} \exp(-2E_{LR}/\hbar\omega) \quad (18)$$

The exponent therein is known as Holstein's reduction factor. The F' electron energy also renormalizes as it is lowered directly through lattice relaxation:

$$E'_{F'} = E_{F'} - E_{LR} \quad (19)$$

We consider a color center system of paramagnetic (unpaired spin) F centers and diamagnetic (paired spins) F' centers in ground state. In a dense color center system there will be electron hopping transitions by tunneling from F' to F centers at energies near $E'_{F'}$ forming a polaron band, to be referred to as *Anderson's band*. Namely, this narrow band is centered at energy $E'_{F'}$ from eq. (19) with a half width $t'_{F'}$ from eq. (18). Transitions in Anderson's band are equivalent to effective F' migration over the color center manifold.

For the alkali halide hosts listed in Tables 1&2 Holstein's reduction factors are from $10^{-8}$ (RbF) to $10^{-33}$ (NaI). These brief estimates suggest that hopping in Anderson's band is too meager to affect in any way the migration pattern in the color center system. Under these conditions F' centers will stay localized and only influence the optical spectra of the host crystals. Our conclusions at this point are similar to analyses elsewhere.[11] However at high local densities of F, F' and mobile α centers, Anderson-type tunneling transitions would lead to the formation of F center clusters which control the color center photo-aggregation process.[1]

## 4. Conclusion

It seems safe to conclude that the negative-U model is in concert with the observed optical absorption spectra and peak energies of the F' centers in NaI. We believe to have shown that for this host crystal, and perhaps for the remaining iodides too, the lattice relaxation effects are superior in promoting the local pairing of electrons at F' centers as opposed to the trapping of two electrons in the sole field of the anion vacancy.

Table 1
Semicontinuum parameters for F centers[a]

| Host xtal | Cation-Anion Separation $r_0 = a$ (Å) | $r_+/r_-$ | Madelung Energy $V_M$ (eV) | Dielectric Constants $\varepsilon_0$ | $\varepsilon_s$ | Phonon Frequencies $\omega_{Fg}$ $(10^{13}s^{-1})$ | $\omega_{LO}$ | Force Constants $K = M\omega_{g/LO}^2$ (eV/Å$^2$) | |
|---|---|---|---|---|---|---|---|---|---|
| CsI | 3.956 | 0.798 | 6.37 | 2.62 | 6.59 | | 1.79 | | 26.47 |
| RbI | 3.671 | 0.669 | 6.79 | 2.59 | 4.91 | 1.04 | 2.04 | 5.75 | 22.11 |
| KI | 3.533 | 0.606 | 7.06 | 2.62 | 5.10 | 1.60 | 2.70 | 6.22 | 17.72 |
| NaI | 3.237 | 0.471 | 7.73 | 2.93 | 7.28 | 1.96[b] | 3.47 | 5.50 | 17.21 |
| NaI | | | | | | 2.66[c] | | 10.16 | |
| NaI | | | | | | 2.15[d] | | 6.63 | |
| LiI | 3.000 | 0.364 | 8.19 | 3.80 | 16.85 | | | | |
| CsBr | 3.720 | 0.898 | 6.76 | 2.42 | 6.67 | | 1.98 | | 32.39 |
| RbBr | 3.445 | 0.758 | 7.26 | 2.34 | 4.86 | 1.35 | 2.45 | 9.68 | 31.89 |
| KBr | 3.298 | 0.683 | 7.58 | 2.34 | 4.90 | 1.76 | 3.21 | 7.53 | 25.05 |
| NaBr | 2.989 | 0.525 | 8.37 | 2.59 | 6.28 | 3.77[c] | 3.98 | 20.41 | 22.64 |
| NaBr | | | | | | 2.60[d] | | 9.72 | |
| LiBr | 2.751 | 0.404 | 9.02 | 3.17 | 13.25 | | | | |
| CsCl | 3.571 | 0.973 | 7.05 | 2.62 | 7.20 | | 3.1 | | 79.40 |
| RbCl | 3.291 | 0.818 | 7.64 | 2.19 | 4.92 | 1.54 | 3.4 | 12.60 | 61.42 |
| KCl | 3.147 | 0.739 | 7.94 | 2.19 | 4.84 | 1.86 | 4.02 | 8.41 | 39.28 |
| NaCl | 2.820 | 0.558 | 8.86 | 2.34 | 5.90 | 2.77 | 5.1 | 10.97 | 37.17 |
| LiCl | 2.570 | 0.420 | 9.68 | 2.78 | 11.95 | | 7.5 | | 24.26 |
| CsF | 3.004 | 1.259 | 8.29 | 2.16 | | | | | |
| RbF | 2.815 | 1.117 | 8.81 | 1.96 | 6.48 | 2.47 | 5.4 | 32.42 | 154.93 |
| KF | 2.674 | 1.011 | 9.33 | 1.85 | 5.46 | 2.78 | 6.1 | 18.79 | 90.45 |
| NaF | 2.317 | 0.742 | 10.77 | 1.74 | 5.05 | 5.10 | 8.1 | 37.17 | 93.77 |
| LiF | 2.004 | 0.507 | 12.37 | 1.96 | 9.01 | 5.86 | 12.6 | 14.81 | 68.48 |

[a]Cation-anion separations, Madelung energies and dielectric constants are reproduced from Fowler's book, Appendices A & B. Physics of Color Centers, W.B. Fowler, ed. (Academic, New York, 1968). [b]Raman data by D.S. Pan and F. Luty, in *Light Scattering in Solids*, M. Balkanski, ed. (Flammarion, Paris, 1976), p. 539. [c]Fitting F-F' conversion data by Georgiev & Mladenov, J. Phys. Chem. Solids **47** (8) 815-824 (1986). [d]Raman data by F. De Matteis, M. Leblans and D. Schoemaker, Phys. Rev. B **49** (14) 9357-9364 (1994).

Table 2

Negative-U parameters for F' Centers[e]

| Host xtal | Lattice Relaxation Energy $E_{LR}=(V_M/a)^2/2K_{g\,/\,LO}$ (eV) | | Correlation Energy $U=e^2/2\varepsilon_0 a$ (eV) | Negative-U Potential $V_0=V_M-\chi+4E_{LR}-U$ (eV) | | Range Borders $(2a/\pi\hbar)\surd(2mV_0) = (2/\pi)R$ I(0÷1)II(1÷3)III(3-) (Å) | | F' Band Energies (eV) $E_{F'edge}$ | $E_{F'peak}$ |
|---|---|---|---|---|---|---|---|---|---|
| CsI | | 0.049 | 0.694 | | 5.372 | | 2.99048 | 0.55 | 1.48 |
| RbI | 0.297 | 0.077 | 0.756 | 6.722 | 5.842 | 3.10421 | 2.89389 | 0.55 | 1.2 |
| KI | 0.321 | 0.113 | 0.777 | 6.085 | 6.235 | 2.84244 | 2.87726 | 0.70 | 1.60 |
| NaI[b] | 0.518 | 0.166 | 0.758 | 8.544 | 7.136 | 3.08596 | 2.82025 | 1.9 | 2.44 |
| NaI[c] | 0.281 | | | 7..596 | | 2.90973 | | | |
| NaI[d] | 0.430 | | | 8.192 | | 3.02172 | | | |
| LiI | | | 0.631 | | | | | | |
| CsBr | | 0.051 | 0.799 | | 5.665 | | 2.88775 | 0.54 | 1.20 |
| RbBr | 0.229 | 0.070 | 0.892 | 6.784 | 6.148 | 2.92650 | 2.78595 | 0.4 | 0.85 |
| KBr | 0.351 | 0.105 | 0.932 | 7.552 | 6.568 | 2.95596 | 2.75667 | 0.65 | 1.25 |
| NaBr[c] | 0.192 | 0.173 | 0.929 | 7.709 | 7.633 | 2.70671 | 2.47888 | 1.55 | 2.45 |
| NaBr[d] | 0.403 | | | 8.553 | | 2.85103 | | | |
| LiBr | | | 0.824 | | | | | | |
| CsCl | | 0.0025 | 0.769 | | 5.881 | | 2.88244 | | |
| RbCl | 0.214 | 0.044 | 0.998 | 6.998 | 6.318 | 2.83943 | 2.69795 | 0.64 | 1.08 |
| KCl | 0.378 | 0.081 | 1.043 | 7.909 | 6.621 | 2.88652 | 2.64104 | 0.90 | 1.60 |
| NaCl | 0.450 | 0.133 | 1.090 | 9.070 | 7.802 | 2.76994 | 2.56903 | 1.72 | 2.55 |
| LiCl | | 0.292 | 1.006 | | 9.342 | | 2.56195 | | 3.70 |
| CsF | | | 1.831 | | | | | 0.78 | 1.34 |
| RbF | 0.151 | 0.032 | 1.303 | 7.611 | 7.135 | 2.53289 | 2.45241 | | |
| KF | 0.324 | 0.067 | 1.453 | 8.673 | 7.645 | 2.56840 | 2.41139 | 1.24 | 1.77 |
| NaF | 0.291 | 0.115 | 1.783 | 9.651 | 8.947 | 2.34763 | 2.26038 | 1.85 | 2.64 |
| LiF | 1.286 | 0.278 | 1.831 | 15.183 | 11.151 | 2.54679 | 2.18259 | | 2.0 |

[e]F' band optical data from Ref. [1].

For [b,c,d] see the footnotes to Table 1.

## Table 3

F' center bound state energies $E'_n$ (eV)

| Host xtal | Root $x_g$ | Root $x_{LO}$ | Root $y_g$ | Root $y_{LO}$ | $E'_{ng}$ | $E'_{nLO}$ |
|---|---|---|---|---|---|---|
| CsI | | 4.295871 | | -1.900504 | | -0.879 |
| | | 4.696842 | | -0.073028 | | -0.001 |
| RbI | 4.165880 | | -2.534155 | | -1.816 | |
| | 4.837832 | | 0.610076 | | -0.105 | |
| NaI[b] | 4.179879 | | -2.463176 | | -2.206 | |
| | 4.822833 | | 0.534829 | | -0.104 | |
| NaI[d] | 4.251874 | | -2.109314 | | -1.618 | |
| | 4..743839 | | 0.149243 | | -0.008 | |
| NaI | | 2.532999 | | 3.635044 | | -4.805 |
| KBr | 4.363366 | | -1.587923 | | -0.883 | |
| | 4.625847 | | -0.401332 | | -0.056 | |
| NaBr[c] | -2.51 | | -3.431039 | | -5.020 | |

For [b,c,d] see the footnotes to Table 1.